\documentclass[12pt]{iopart}
\usepackage{graphicx}
\usepackage{bm}
\usepackage{color}
\begin{document}

\title{Critical exponents of the pair contact process with diffusion}
\author{R. D. Schram$^1$ and G. T. Barkema$^{1,2}$}
\address{$^1$ Instituut-Lorentz, Leiden University, 
  P.O. Box 9506, 2300 RA  Leiden, The Netherlands}
\address{$^2$ Institute for Theoretical Physics, Utrecht University, 
  P.O. Box 80195, 3508 TD  Utrecht, The Netherlands}
\eads{raouldschram@gmail.com}

\begin{abstract}
We study the pair contact process with diffusion (PCPD) using Monte
Carlo simulations, and concentrate on the decay of the particle density $\rho$
with time, near its critical point, which is assumed to follow
$\rho(t) \approx ct^{-\delta} +c_2t^{-\delta_2}+\dots$.
This model is known for its slow convergence to the
asymptotic critical behavior; we therefore pay particular attention to
finite-time corrections. We find that at the critical point, the ratio
of $\rho$ and the pair density $\rho_p$ converges to a constant, 
indicating that both densities decay with the same powerlaw.
We show that under the assumption $\delta_2 \approx 2 \delta$, two of the
critical exponents of the PCPD model are
$\delta = 0.165(10)$ and $\beta = 0.31(4)$,
consistent with those of the directed percolation (DP) model.
\end{abstract}

\maketitle

\section{Introduction}

The pair contact process with diffusion (PCPD) is a one-dimensional
model of fermionic particles on a lattice. In this context, {\it fermionic} means that
a site cannot be occupied by more than one particle. When two particles are
adjacent, they can interact in two ways: they can annihilate each other,
or alternatively, they can create another particle on an adjacent
lattice site.  Particles can also diffuse by hopping from one site to
the next. The reactions of the PCPD model and their rates are given by:
\begin{eqnarray}
\begin{array}{ccc}
\left\{
\begin{array}{ccc}
AA0 & \rightarrow & AAA \\
0AA & \rightarrow & AAA
\end{array} \right.
&{\rm each\,\,with\,\, rate}&
\frac{(1-p)(1-d)} 2 \label{2Ato3A} \\
AA~ \rightarrow ~ 00
&{\rm with \,\,rate}& p\,(1-d) \label{2Ato0}\\
A0  \leftrightarrow  0A &{\rm with \,\,rate}&  d \label{A0to0A}
\end{array}
\label{eq:rates}
\end{eqnarray}

Given a value for the diffusion coefficient $d$, we can discern three
different regimes depending on the annihilation rate $p$. If $p$ is
very large, then annihilation dominates the process, and the particles
on the lattice will die out quickly. This is the inactive phase. On the
other hand, if $p$ is very small, the particle creation reaction will
ensure that (with extremely high probability) the system will maintain
a high density. This is called the active phase. Well into the active or
the inactive phase, long-ranged interactions are absent, and both these
regimes can therefore be described well by mean-field theory.
However, if the boundary between the active and inactive phase is approached
from either side, length and time scales diverge in a power-law fashion.
In analogy with equilibrium phase transitions, one expects that the system
then exhibits critical behaviour in the transition between these regimes, at the
critical value $p_c$. From equilibrium statistical physics it is
well established that phase transitions can be classified in universality
classes, each of which characterized by a unique set of critical exponents.
Moreover, these exponents are typically insensitive to small changes
in the model, such as details of short-ranged interactions. 
A central question in non-equilibrium statistical physics is
whether also dynamical phase transitions can be classified into universality
classes, which also are insensitive to such small changes.  

In this paper we will concentrate on the scaling relations concerning
the particle density $\rho$, which are given by:

\begin{eqnarray}
\begin{array}{rclc}
\rho_{p=p_c} & \sim & t^{-\delta} & (\epsilon = 0) \\
\rho_{t\to \infty} & \sim & \epsilon^\beta  & (\epsilon > 0)\\
\end{array},
\end{eqnarray}

where $\epsilon \equiv \left| p-p_c\right|$ is the distance from criticality,
and $\delta$ and $\beta$ are two of the critical exponents of the PCPD
system. A conjecture by Grassberger \cite{grass} and Janssen
\cite{janssen} states that all systems with a single order parameter and a single
absorbing state will belong to the universality class of the Directed
Percolation (DP) model. The critical exponents of the one-dimensional DP model are known
to very high accuracy ($\delta = 0.159464(6)$ and $\beta=0.276486(8)$) \cite{jensen99}.
The values for $\delta$ and $\beta$ of the PCPD model have been disputed extensively, with estimates of $\delta$ ranging from $0.16(1)$ to $0.27(4)$ and estimates of $\beta$ varying from $\beta<0.34$ to $\beta=0.58(1)$ \cite{small,odor00,carlon01,hinrichsen01,park01,park02,dickman02,odor03,kockelkoren,barkema03,nohpark,park03,hinrichsen,oliveira,kwon}. For a more extensive overview,
see \cite{small}. The main interest in the PCPD model is whether it
shares its critical exponents with DP, or not; the latter would disprove the
Grassberger-Jansen conjecture.

The difficulties with the determination of the exponent $\delta$
of the PCPD model is that there are large finite-time corrections
\cite{small,hinrichsen}, and the effective exponent thus shows a drift
with simulation time. In this paper we generate high-quality data,
exploiting the computational power of graphics processing units (GPUs); a description of
the implementation will be published elsewhere. We
then analyze these data in a way that suppresses the leading finite-time
corrections. 

\section{Simulation results}

At criticality, the asymptotic decay of the density is described by
a power-law of the form $\rho(t) \sim t^{-\delta}$. For the DP
model the corrections to this power-law decay very rapidly, and
therefore $\delta_{\rm DP}$ is known with very high accuracy. However,
as the history of the PCPD model shows, the density decay in the PCPD
model does not show such a clean power-law in the time range accessible
to computer simulations.  Consequently, one must account for finite-time
corrections, in order to obtain an accurate estimate of $\delta$.
Different finite-time corrections have been proposed in the past,
including logarithmic corrections \cite{odor03} and power-law corrections
\cite{barkema03}. Our data suggest power-law corrections, so that
the density decay at the critical point is given by

\begin{equation} \label{eq:cor}
 \rho(t) = c_1 t^{-\delta} + c_2 t^{-\delta_2} + \ldots, 
\end{equation}

with $\delta<\delta_2<\ldots$. By differentiating the logarithm
of the density vs. the logarithm of the time, one can define an effective
decay exponent as

\begin{equation} \label{eq:eff}
\delta_{\rm eff}(t) \equiv -\frac{\partial \log(\rho)}{\partial \log(t)}
  \approx \delta + \frac{c_2}{c_1} (\delta_2-\delta) t^{-(\delta_2-\delta)} +\ldots.
\end{equation}

Note that in the limit of infinite time, the effective exponent $\delta_{\rm eff}$ 
approaches the true asymptotic value $\delta$. Moreover, the direction from which it
approaches the asymptotic value is governed by the sign of $c_2/c_1$, and the speed
of convergence by the gap $\delta_2-\delta$. 

At this point, we want to make some practical remarks: 

i) In practice with simulation data, the differentiation is carried out
numerically, and one makes the approximation $\delta_{\rm eff} \approx
\frac{-\log (\rho(2t)/\rho(t))}{\log 2}$.

ii) When simulation measurements of $\delta_{\rm eff}$ are presented, it
is usually more convenient for the eye if the presentation is such that
the infinite-time behavior is inside the plot, rather than at infinite
distance. A common approach to achieve this, is to plot $\delta_{\rm eff}$
as a function of $1/t$; this is not a good idea! If the gap is
less than 1, the approach to the vertical axis will eventually become infinitely
steep, and it is hard to predict how far an infinitely steep curve will
shoot up (or down).  Ideally, one should therefore plot $\delta_{\rm eff}$
as a function of $t^{-(\delta_2-\delta)}$, as then the approach to the
vertical axis will follow a straight line (and indirectly provide information
on $\delta_2-\delta$).  Alternatively, one can plot $\delta_{\rm eff}$ as a
function of $\rho^{(\delta_2/\delta-1)}$.

iii) Besides the particle density $\rho$, another numerically accessible
quantity is the pair density $\rho_p$, defined as the fraction of
neighboring sites, both of which are being occupied by a particle. For
the pair density at the critical point, relations equivalent to
eqs.(\ref{eq:cor}) and (\ref{eq:eff}) can be defined. Barkema and
Carlon~\cite{barkema03} provided numerical evidence that the ratio of the
densities of single particles and pairs tends to a constant; further on
in this paper we will confirm this. This has as a consequence that the
asymptotic density decay of singles and pairs is governed by a unique
exponent $\delta$. Also, this means that as long as the effective exponents
$\delta_{\rm eff}$ and $\delta_{p, {\rm eff}}$ do not coincide, 
finite-time corrections are still significant.

\begin{figure}\centering
\includegraphics[width=10cm]{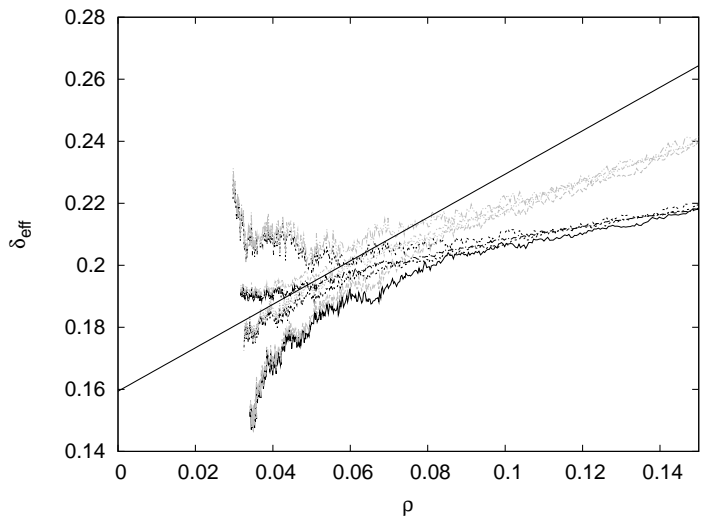}
\caption{The effective exponent $\delta_{\rm eff}$ as obtained for both
the particle density $\rho$ (colored black) and the pair density $\rho_p$
(gray), as a function of the particle density, with a diffusion factor $d=0.5$. For $d=0.25$ and $d=0.75$, we found similar results. Starting from low to high 
in the picture, the simulations are performed with $p=0.15246$
(1600 runs), 0.15247 (3300 runs), 0.152475 (7400 runs) and 0.152485 (1000
runs). The lattice contained $L=2^{18}=262144$ sites. Each simulation
reached a time of approximately $t=1.6 \cdot 10^{7}$.  The curve for
$p=0.152475$ is closest to the critical value, which we estimate to be
$p_c = 0.152473(2)$. The straight line signifies a possible extrapolation
to $\delta=\delta_{\rm DP}$.}
\label{fig:delta_eff}
\end{figure}

The first results, presented in figure \ref{fig:delta_eff}, are
measurements of $\delta_{\rm eff}$ as a function of $\rho$, for values
of $p$ close to the critical point, on either side.  The figure
shows that $\delta_{\rm eff}$ approaches its asymptotic value from
above.  Assuming that at the end of our simulations, corrections
are dominated by $\delta_2$, this indicates that $c_2$ is positive.  
Furthermore, under this assumption, the maximal value of
$\delta_{\rm eff}$ with a $p$ value lying in the inactive phase provides an upper bound for the asymptotic exponent:
$\delta < 0.19$ at $p=0.15247$. This bound is less tight then the bound reported by
Hinrichsen\cite{hinrichsen}, because it depends strongly on the maximal
simulation time of the dataset, which was higher in Hinrichsen's
simulation, and less on the accuracy of the dataset, which is higher in
our simulations. We do however confirm the conclusion of Hinrichsen
that most of the values reported for the exponent $\delta$ in PCPD
are ruled out by this upper bound. 

Interestingly, the data in figure \ref{fig:delta_eff} are consistent with
a linear convergence to an asymptotic value of $\delta=\delta_{DP}$.
The most important conclusion drawn from this observation is that our
data do not provide evidence of a violation of the Grassberger-Janssen
conjecture; and since the product of system size, length of the simulations, and the number of these is higher in our simulations than in earlier studies which claim to provide evidence for violation of the conjecture, our conclusion is that earlier claims of such violation based
on numerical measurements of $\delta_{\rm eff}$ are ill-founded.  A second 
observation is that the observed linear convergence indicates that
$\delta_2/\delta-1 \approx 1$ and thus that $\delta_2 \approx 2\delta$.
Although our simulations are more extensive than those of earlier reports,
the data is still not accurate enough to unambiguously rule out combinations
of $(\delta,\delta_2)$ which are slightly different from $(\delta_{DP},2\delta_{DP})$;
in particular there is quite some ambiguity in $\delta_2$.

%%%%%% FIG %%%%%%

We now turn to an accurate estimation of the critical point $p_c$.
For this, we want to capitalize on our numerical knowledge on the
value for the correction exponent $\delta_2$, to suppress finite-time
corrections. We do so by defining an adjusted density:

\begin{equation} \label{eq:rho_cor}
 \rho^*(t) = 2 \rho(t) - \rho(2^{-1/\delta_2} t),
\end{equation}

for some assumed value of $\delta_2$. Asymptotically, the densities
$\rho(t)$ and $\rho^*(t)$ at the critical point will coincide, but at
long but not infinite times, the corrected density suppresses correction
terms with exponents $\delta_i$ close to $\delta_2$; with the correct
numerical choice for $\delta_2$ the first correction term will even be
completely removed, and for a reasonable approximation of $\delta_2$, the
power-law decay of $\rho^*(t)$ should be much cleaner than that of $\rho(t)$.

\begin{figure}\centering
\includegraphics[width=10cm]{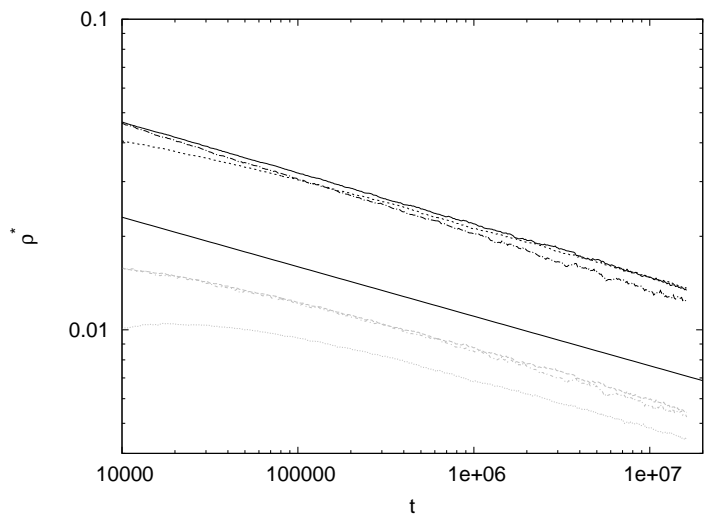}
\caption{Corrected particle density $\rho^*$
(black) and corrected pair density $\rho^*_p$ (gray) as a function of time,
for $d=0.25$, $p=0.125142$ (dashed); $d=0.5$, $p=0.152475$ (solid); and
$d=0.75$, $p=0.191790$ (dotted-dashed). In the correction procedure, we used
$\delta_2=0.319$ and a lattice size of $L=2^{18}=262144$. The thick black line shows a
possible line with $\delta=\delta_{\rm DP}$.}
\label{fig:cor}
\end{figure}

Figure \ref{fig:cor} shows the corrected density and corrected pair
density as a function of time in a log-log plot, with the choice
$\delta_2=2 \delta_{\rm DP} = 0.319$. This gives us the estimate
$\delta=0.165(10)$. We conclude that the choice of $\delta_2$ tightly
defines our estimates of both $\delta$ and $p_c$.

%%%%%% FIG %%%%%%

An accurate method to determine the critical point $p_c$ is to use
a data collapse method on the corrected density $\rho^*$. The data collapse method
maps the dataset with off-critical $p$ values onto a single curve for $p>p_c$ and
a second curve for $p<p_c$, if we select the correct values for $\delta$, $\beta$
and $p_c$. To achieve this, the (pair) density and the time are rescaled
as follows:

\begin{equation} \label{eq:collapse}
 t' = \epsilon^{\beta/\delta} t, \\
 \rho' = \epsilon^{-\beta} \rho^*.
\end{equation}

\begin{figure}\centering
\includegraphics[width=10cm]{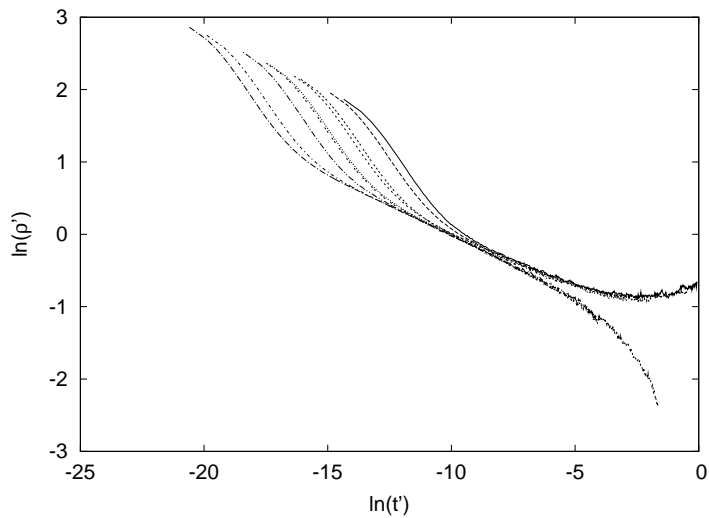}
\caption{Data collapse of curves obtained with various values for $p$ at $d=0.5$.
The rescaled corrected density is plotted as a function of rescaled
time (see Eqs.~(\ref{eq:collapse})). The data are obtained from the
same simulations as in Fig.~\ref{fig:delta_eff}; added to these data
are simulations with $p=0.1524$ (500 runs on a smaller lattice with
$L=65536$), $p=0.15242$ (1000 runs), $p=0.15245$ (1000 runs), $p=0.15248$
(500 runs) and $p=0.1525$ (1000 runs). In the correction procedure, we
used $\delta_2=0.319$. In the rescaling, we used  $\delta=0.159$ and
$\beta=0.2765$ as in the DP model, combined with $p_c=0.152473$.}
  \label{fig:collapse}
\end{figure}

Using the diffusion factor $d=0.5$, we obtain a very good data collapse,
shown in figure \ref{fig:collapse}. We find that the critical
value is $p_c = 0.152473(2)$. Our estimate of the exponent $\beta$ is
less precise then our estimate of $\delta$: $\beta=0.31(4)$, which is
consistent with the value known for DP: $\beta=0.2765$. The fact that 
the curve for $p=0.15240$ shows a statistically significant increase,
after reaching a (pseudo-)equilibrium, indicates that the corrections
due to finite time and due to off-critical values for $p$ are not simply
additive: our procedure to suppress the leading finite-time corrections at
$p_c$ is not working equally well for the off-critical curves. The `best'
value we find for $\beta$ is therefore still experiencing significant
corrections to scaling.

%%%%%% FIG %%%%%%

The question that remains is how these parameters depend on our choice for
$\delta_2$. We used the same methods described above to obtain estimates
with $\delta_2$ ranging from 0.319 to 0.5. The results are presented in
Table \ref{tbl:d2}. Note that for our choice $\delta_2=0.319$, the leading
exponent $\delta$ shows little dependence on the diffusion coefficient $d$.

\begin{table}\centering
\caption{Values for the exponents $\delta$ and $\beta$ and critical
point $p_c$, obtained by our analysis approach, for three values of the
diffusion coefficient $d$ and for our estimated value $\delta_2=0.319$,
as well as for some other values for $\delta_2$.
\label{tbl:d2}
}
\begin{tabular}{lrrrr}
$d$ & $\delta_2$ & $\delta$ & $\beta$ & $p_c$ \\
\hline

$0.25$ & $0.319$ & $0.176(8)$ & - & $0.125141(2)$ \\
$0.5$ & $0.319$ & $0.164(4) $ & $0.31(4)$ & $0.152473(2)$ \\
$0.75$ & $0.319$ & $0.159(4)$ & - & $0.191789(2)$ \\
\hline
$0.25$ & $0.37$ & $0.187(8)$ & - & - \\
$0.5$ & $0.37$ & $0.175(4)$ & $0.32(4)$ & $0.152476(2)$ \\
$0.75$ & $0.37$ & $0.165(4)$ & - & - \\
\hline
$0.25$ & $0.43$ & $0.190(8)$ & - & - \\
$0.5$ & $0.43$ & $0.181(4)$ & $0.32(4)$ & $0.152476(2)$ \\
$0.75$ & $0.43$ & $0.172(4)$ & - & - \\
\hline

$0.25$ & $0.5$ & $0.195(8)$ & - & - \\
$0.5$ & $0.5$ & $0.185(4)$ & $0.33(4)$ & $0.152477(2)$ \\
$0.75$ & $0.5$ & $0.177(4)$ & - & - \\
\hline

\end{tabular}
\end{table}

In Ref.~\cite{small} the leading correction term $\sim t^{-\delta_2}$ was removed
by using a linear combination of the density $\rho$ and the pair density
$\rho^*$. This is a valid strategy as long as the dominant correction term
($c_2 t^{-\delta_2}$) does not cancel in the ratio $\rho/\rho_p$. In that
case this ratio will asymptotically go to a constant with a power law
(at criticality):

\begin{equation}
 \rho/\rho_p = k_1 + k_2 t^{\delta-\delta_2} + \ldots
\end{equation}

\begin{figure}
\includegraphics[height=6cm]{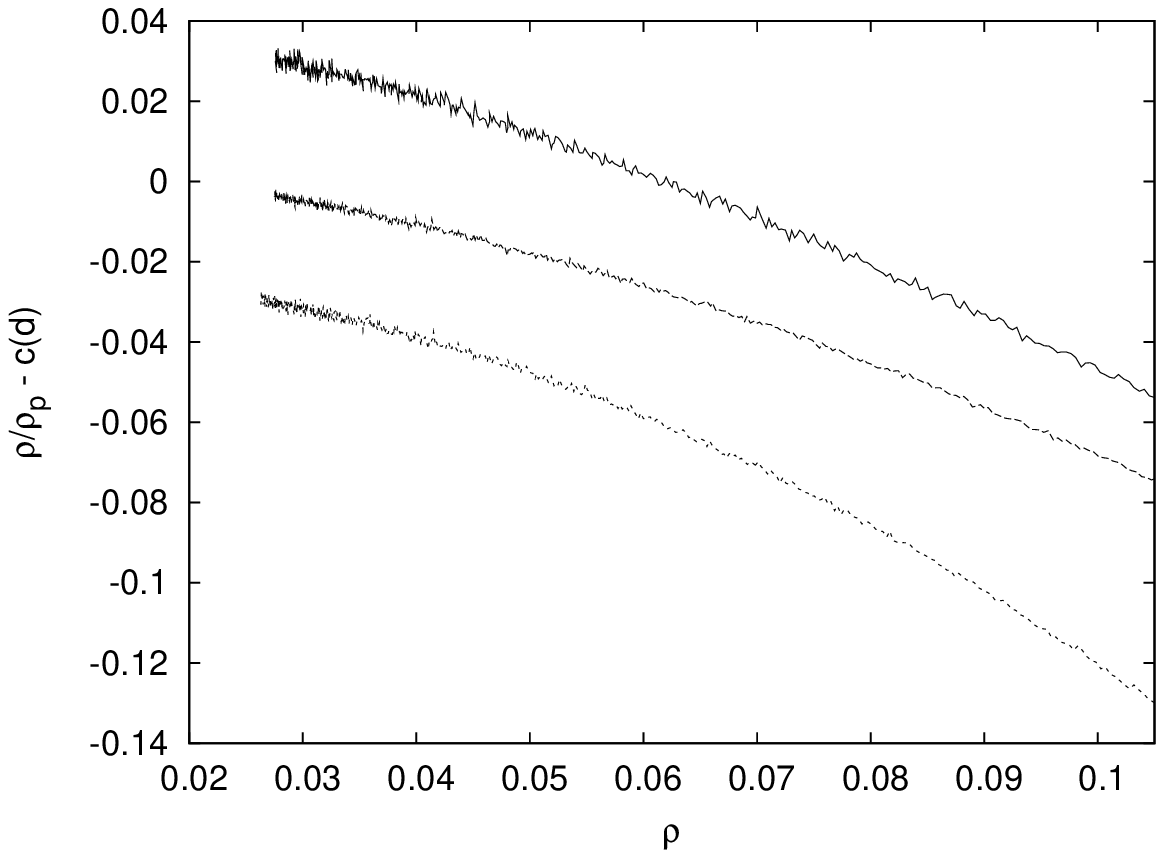}
\includegraphics[height=6cm]{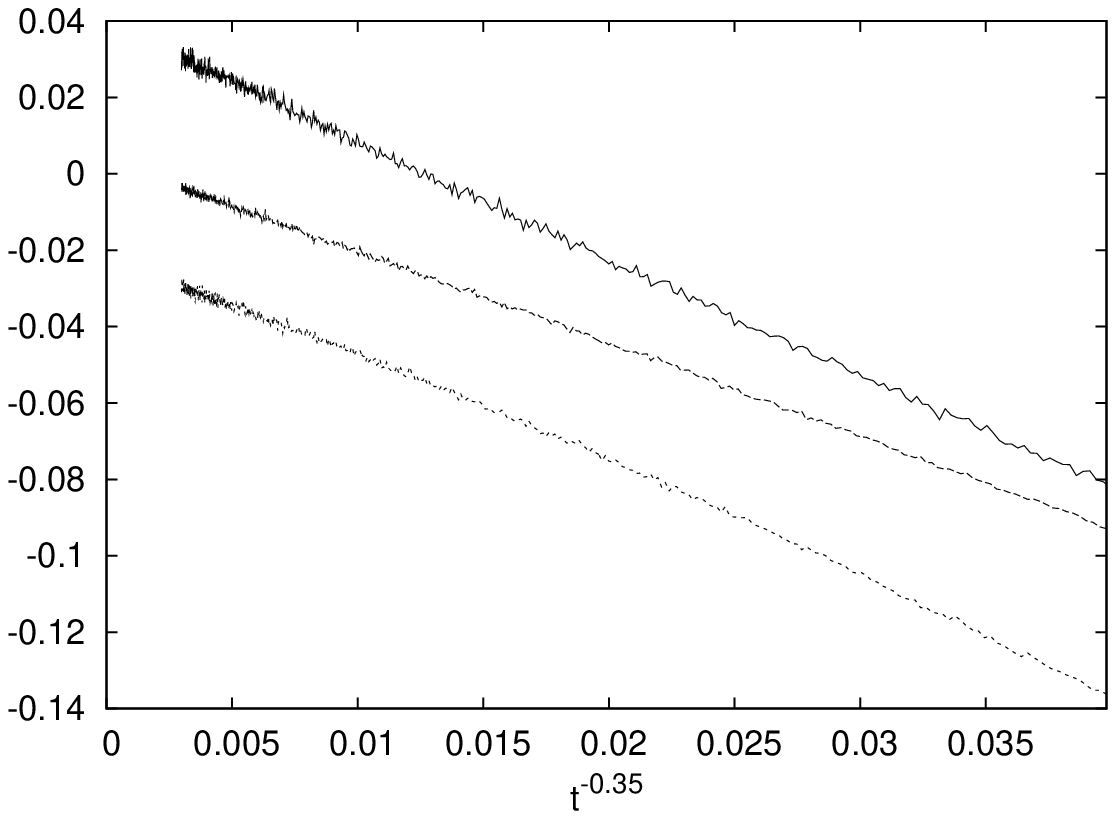}
\caption{Left panel: ratio $\rho/\rho_p$ of the particle and pair densities,
as a function of the particle density. Right panel: same data, plotted as
a function of $t^{\delta-\Delta}$ with $\Delta=0.51$. The different curves
correspond to $d=0.25$ (solid line), $d=0.5$ (larger
dashes) and $d=0.75$ (smaller dashes). Curves are shifted vertically by some
arbitrary, $d$-dependent constant $c(d)$, to let them fit nicely into
a single figure; lower curves correspond to higher values of $d$.}
\label{fig:rat}
\end{figure}

Thus, if the density ratio $\rho/\rho_p$ is plotted against $t^{\delta-\delta_2}$, 
one should find a straight line (with finite-time corrections
from higher-order correction terms). Since we found previously that
$\delta_2 \approx 2 \delta$, we have plotted the ratio versus the density
$\rho$ in figure \ref{fig:rat}. The density ratios for the
three values of $d$ are clearly not approaching the vertical axis under a fixed angle in this
plot, but seem to arrive horizontally at $\rho=0$. Thus, we find that the
leading correction disappears in the ratio and the method described in
\cite{small} fails. 

We plotted the ratio against $t^{\delta-\Delta}$ and we found that the
most straight line has a value of $\Delta \approx 0.51$. The large
discrepancy between $\delta_2$ and $\Delta$ increases our confidence
in the fact that the correction term $\sim t^{-\delta_2}$ vanishes in
the ratio $\rho/\rho_p$.

%%%%%% FIG %%%%%%

Thus, the ratio between the particle density and the pair density goes to
a constant even faster than expected. This still means that the exponent
$\delta$ is equal for the particle and the pair density.

\section{Summary and conclusion}

We have performed extensive simulations of the one-dimensional PCPD model,
using a highly efficient GPU-based simulation approach.
The analysis of the simulation results was performed in a way that takes
account of finite-time effects. Our main goal was to verify
the Grassberger-Janssen conjecture, which predicts that the exponents for
PCPD coincide with those of directed percolation. 

We find that our data are consistent with DP values for the exponents
$\delta$ which describes the decay of the particle density at the critical
point, and $\beta$ which describes the asymptotic particle density for
simulations close to the critical point, but slightly in the active phase.

Additionally, we find that the 
the leading correction exponent $\delta_2$ is numerically found to be close
to $2\delta$, which would suggest corrections to scaling of order $\rho^2$.
We also find that these leading corrections do not seem to manifest themselves
in the ratio of the particle and pair densities.

Under the assumption that the Grassberger-Janssen conjecture is correct, we find strong indications that $\delta_2 \approx 2 \delta$, by using the corrected density. There is a small dependence of $\delta$ on the diffusion factor $d$. We attribute this to the differences in the amplitudes of the second- and higher-order corrections. This induces a (small) systematic error in the values for $\delta$ given in Table 1.
\ack

Computing time on the ``Little Green Machine", which is funded by the Dutch
agency NWO, is acknowledged.

\end{document}